\documentclass[aps, prl, 10pt, amsmath, amssymb, floatfix, superscriptaddress, twocolumn]{revtex4-2}
\usepackage{physics,esint,graphicx,bbm,amsthm}
\usepackage[dvipsnames,svgnames,table]{xcolor}
\usepackage[unicode=true,
  pdfusetitle,
  bookmarks=true,
  bookmarksnumbered=false,
  bookmarksopen=false,
  breaklinks=true,
  pdfborder={0 0 0},
  backref=false,
  colorlinks=true,
hypertexnames=false]{hyperref}
\hypersetup{linkcolor=NavyBlue,urlcolor=NavyBlue,citecolor=NavyBlue}

\newcommand*{\TT}{{\mkern-1.5mu\mathsf{T}}} 
\newcommand{\iqfim}{\mathcal{F}^{-1}} 

\newcommand{\mcE}{\mathcal{E}}
\newcommand{\mcF}{\mathcal{F}}

\newcommand{\mcQ}{\mathcal{Q}}

\begin{document}
\title{Quantum Parameter Estimation Uncertainty Relation}
\author{Bing-Shu Hu}
\author{Xiao-Ming Lu}
\email{lxm@hdu.edu.cn}
\affiliation{School of Sciences and Zhejiang Key Laboratory of Quantum State Control and Optical Field Manipulation, Hangzhou Dianzi University, Hangzhou 310018, China}

\begin{abstract}
  Quantum multiparameter estimation focuses on the simultaneous inference of multiple parameters in quantum systems through measurement and data processing.
  Its complexity stems from two key factors: measurement incompatibility and parameter correlation.
  By strategically manipulating the multidimensional parameter space, we derive an estimation uncertainty relation that quantifies how these factors jointly limit estimation precision in the two-parameter case. 
  This uncertainty relation is tight for pure states and thus completely describes the quantum limit of two-parameter estimation precision in a simple inequality.
  To intuitively illustrate the impact of the uncertainty relation, we develop an error-ellipse method and demonstrate its utility in phase-space displacement estimation.
  Our results reveal that a geometric perspective of the parameter space offers a powerful approach for addressing multiparameter estimation challenges. 
\end{abstract}

\maketitle

Since the pioneering work of Helstrom~\cite{Helstrom1967,Helstrom1968,Helstrom1976}, quantum Cram\'er-Rao bound (QCRB) has become the most popular tool in quantum metrology~\cite{Braunstein1994,Giovannetti2004,Giovannetti2006,Giovannetti2011,Paris2009,Liu2020}.
It has been widely applied in various fields, e.g. optical interferometry~\cite{Demkowicz-Dobrzanski2015}, atomic interferometry~\cite{Pezze2018}, quantum imaging~\cite{Tsang2016b,Tsang2020a} and gravitational wave detection~\cite{Miao2017,Gardner2024}.
Despite its success, the QCRB is not always asymptotically attainable in multiparameter estimation scenarios~\cite{Yuen1973,Matsumoto2005,Ragy2016,Yang2019b,Suzuki2019,Carollo2019a,Kukita2020,Demkowicz-Dobrzanski2020}, making it difficult to assess the fundamental limits of multiparameter estimation precision with quantum systems.
A satisfactory tool for the quantum multiparameter estimation problem is still lacking.

The intricacies of quantum multiparameter estimation stems from two key factors: measurement incompatibility and parameter correlations.
Measurement incompatibility arises because different parameters may require different optimal measurements, leading to a fundamental tradeoff in minimizing estimation errors across parameters~\cite{Lu2021,Belliardo2021,Conlon2021,Goldberg2021a,Albarelli2022,Chen2022,Gill2000,Li2016,Zhu2018}.
Parameter correlation refers to the statistical dependence between parameters, meaning that individual estimation errors alone cannot fully characterize overall precision.
Therefore, parameter correlation further complicates the landscape of multiparameter estimation by introducing additional constraints on the achievable precision.

In this paper, we present a two-parameter estimation uncertainty relation (TEUR) that captures the joint effect of measurement incompatibility and parameter correlation. 
This TEUR concisely bounds the covariance matrix of unbiased estimates by the quantum Fisher information matrix (QFIM) and a scalar we term the incompatibility factor.
The TEUR is tight for pure states and thus completely describes the quantum limit of two-parameter estimation precision for pure states.

\vspace{8pt}
\noindent\textbf{\large Results\\}
Let us consider a general model of quantum multiparameter estimation, which can be described by a parametric density operator $\rho_{\theta}$ depending on an unknown vector parameter $\theta=({\theta_1,\theta_2,\dots,\theta_d})^\TT \in \mathbb{R}^d$ with \(\TT\) denoting the transpose of matrix and vector.
An estimation strategy is constituted by a quantum measurement, described by a positive-operator-valued measure \(\{M_x\}\), and an estimator $\hat{\theta}=(\hat{\theta}_1,\hat{\theta}_2,\dots,\hat{\theta}_d)^\TT$, which maps the observation data $\vec{x} = (x_1,x_2,\ldots,x_n)$ collected from \(n\) samples to the estimates of $\theta$.
The estimation precision is characterized by the covariance matrix $\mcE$, whose entries are $ \mcE_{\mu\nu} = \mathbb{E}[\hat\theta_\mu \hat\theta_\nu] - \mathbb{E}[\hat\theta_\mu] \mathbb{E}[\hat\theta_\nu]$ with $\mathbb{E}[\bullet]$ denoting the expectation with respect to the probability distribution $p_\theta(\vec{x}) = \prod_{j=1}^n \tr(\rho_\theta M_{x_j})$.
The covariance matrix $\mcE$ for any unbiased estimator and any quantum measurement must obey the QCRB: \(\mcE \geq n^{-1} \mcF^{-1}\), where \(\mcF\) is the QFIM~\cite{Helstrom1967,Helstrom1968}.
The elements of the QFIM are defined as $\mathcal{F}_{\mu\nu} = \Re \tr(\rho_{\theta} L_\mu L_\nu)$, where $L_\mu$---the symmetric logarithmic derivative operator---is the Hermitian operator satisfying $\partial_\mu \rho_\theta = (L_\mu\rho_{\theta}+\rho_{\theta}L_\mu)/2$ with \(\partial_\mu\) denoting the partial derivative with respect to \(\theta_\mu\).

We here give a concise tradeoff relation that, in addition to the QCRB, imposes limitations on reducing the errors in estimating two unknown parameters.
Our tradeoff relation---the TEUR---is given by
\begin{equation}\label{eq:Theorema_Egregium}
  \sqrt{|n\mcE\mcF - I|} + \sqrt{(1 - \gamma)|n\mcE\mcF|} \geq 1,
\end{equation}
where \(|\bullet|\) represents the matrix determinant, \(I\) denotes the two-dimensional identity matrix, and \(\gamma\) is the incompatibility factor defined as
\begin{align} \label{eq:gamma}
  \gamma = \frac{\norm{\sqrt{\rho_\theta}[L_1, L_2]\sqrt{\rho_\theta}}_1^2}{4\,|\mathcal F|}
\end{align}
with $\norm{X}_1 \equiv \tr\sqrt{X^{\dagger}X}$ denoting the Schatten-1 norm of an operator \(X\).


The incompatibility factor \(\gamma\) takes values in the range \([0,1]\).
Note that \(\gamma=0\) is equivalent to the partial commutativity condition~\cite{Yang2019b}, which is necessary and sufficient for the QCRB to be saturated.
For the maximum incompatible factor (\(\gamma=1\)), the TEUR becomes \(|n \mcE\mcF - I| \geq 1\).
For diagonal QFIMs, the TEUR with \(\gamma=1\) is equivalent to
\begin{equation} \label{eq:TEUR_extremal}
  \qty(\mcE_{11}-n^{-1}\mcF_{11}^{-1})
  \qty(\mcE_{22}-n^{-1}\mcF_{22}^{-1})
  - \mcE_{12}^2 \geq 1,
\end{equation}
implying that the optimal estimation strategies for one parameter must lead to the divergence of the estimation error of the other parameter.
This extreme case evidently manifests the trade-off between the precision of the two parameters.

For pure states \(\rho_\theta = \op{\psi_\theta}\), the TEUR has some good properties. 
First, it is asymptotically tight for pure states~\cite{SupplementalMaterial}, meaning that for a large number of samples, there always exists a quantum measurement and an unbiased estimator such that the equality in Eq.~\eqref{eq:Theorema_Egregium} holds.
Consequently, the TEUR completely describes the quantum limit of estimation error for two unknown parameters encoded in pure states.
Second, for pure states, both the incompatibility factor and the QFIM can be expressed in terms of the quantum geometric tensor~\cite{Provost1980,Berry1989,Hetenyi2023}
\begin{equation}
  \mcQ_{\mu\nu} = \ip{\partial_\mu\psi_\theta}{\partial_\nu\psi_\theta} - \ip{\partial_\mu\psi_\theta}{\psi_\theta} \ip{\psi_\theta}{\partial_\nu\psi_\theta}
\end{equation}
as \(\mcF = 4 \Re\mcQ\) and \(\gamma = |\Im \mcQ| / |\Re \mcQ|\).
Therefore, the quantum geometric tensor contains the sufficient information about the two-parameter estimation problem with pure states in the asymptotic regime.
This signifies a profound connection between quantum geometry and multiparameter estimation.

\begin{figure}
  \centering
  \includegraphics{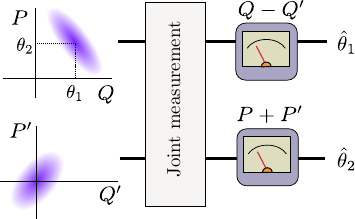}
  \caption{
    Phase-space complex displacement estimation.
    The joint measurement is implemented with an ancillary mode, which is prepared in a squeezed vacuum state.
  }\label{fig:example}
\end{figure}

\vspace{8pt}
\noindent\textbf{Application.} 
We here use the TEUR to study a typical problem in quantum optics---the estimation of a phase-space complex displacement of a bosonic mode.
Denote by \(a\) and \(a^\dagger\) the annihilation operator and creation operator for the bosonic mode, respectively.
Assume that the initial state of the bosonic mode is a squeezed vacuum state \(S(\zeta)\ket{0}\), where \(\ket0\) denotes the vacuum state and \(S(\zeta) = \operatorname{exp}(\zeta^* a^2 / 2 - \zeta a^{\dagger2} / 2 )\) with \(\zeta = r\exp(i\varphi)\) is the squeeze operator.
The state before measurement is a displaced squeezed state \(\ket{\alpha;\zeta} = D(\alpha)S(\zeta)\ket{0}\), where \(D(\alpha) = \exp(\alpha a^\dagger - \alpha^* a)\) with \(\alpha\) being an unknown complex number.
The parameters of interest are the real and imaginary parts of $\alpha$ and denoted by $\theta_1=\Re\alpha$ and $\theta_2=\Im\alpha$.
For this model, we obtain the quantum geometric tensor~\cite{SupplementalMaterial}:
\begin{align} \label{eq:Q}
  \mathcal{Q} = R\qty(\frac{\varphi}{2}) C(2r) R\qty(\frac{\varphi}{2})^\TT + \mqty(0 & i \\ -i & 0),
\end{align}
where
\begin{align}
  R(\varphi) = \mqty(\cos\varphi & - \sin\varphi \\ \sin\varphi & \cos\varphi ),
  \quad
  C(r) = \mqty(\dmat[0]{e^r, e^{-r}}).
\end{align}
As the incompatibility factor for this model is \(\gamma = 1\), the TEUR becomes \(|n \mathcal{E}\mathcal{F}  - I |\geq 1\).

The TEUR can be saturated by the following measurements.
We first prepare an ancillary bosonic mode and then measure a pair of commuting observables \(\mathcal A = Q - Q'\) and \(\mathcal B = P + P'\), where \(Q  = (a+a^\dagger) / 2\) and  \(P=(a-a^\dagger)/(2i)\) are the quadrature operators for the system and \(Q'\) and \(P'\) are the quadrature operators for the ancillary mode, as illustrated in Fig.~\ref{fig:example}.
We take the sample means of outcomes of measuring \(\mathcal A\) and \(\mathcal B\) as the estimate of \(\theta_1\) and \(\theta_2\), respectively.
The estimators are unbiased when the initial state of the ancillary mode satisfies \(\ev{Q'} = \ev{P'}=0\).
As the system and the ancilla are initially uncorrelated, the covariance matrix of estimators is given by
\begin{equation} \label{eq:example_error}
  n \mcE = V(Q,P) +  V(-Q',P'),
\end{equation}
where \(V(X_1,X_2)_{\mu\nu}=\Re\ev{X_\mu X_\nu}-\ev{X_\mu}\ev{X_\nu}\) is defined for any two Hermitian operators \(X_1\) and \(X_2\).
After some algebra, we obtain~\cite{SupplementalMaterial}
\begin{equation}
  V(Q,P) = \frac14 R \qty(\frac{\varphi}{2}) C(-2r) R\qty(\frac{\varphi}{2})^\TT,
\end{equation}
which equals to \(\mathcal F^{-1}\).
Therefore, the second term of Eq.~\eqref{eq:example_error} is the extra error covariance above the QCRB due to measurement incompatibility.
As a result, the TEUR is equivalent to \(|V(-Q',P')|\geq 1/16\) and can be saturated when we prepare the ancillary mode in any minimum uncertainty state, e.g., squeezed vacuum states.

\vspace{8pt}
\noindent\textbf{Error visualization.}
We develop an error-ellipse approach to intuitively manifest the impact of the TEUR on the estimation precision.
For a covariance matrix \(\mcE\), the error ellipsoid is defined as the set of points \(v\) in the parameter space such that 
\begin{equation} \label{eq:error_ellipsoid}
  (v - \bar\theta)^\TT (n \mcE)^{-1} (v - \bar\theta) \leq \kappa^2,
\end{equation}
where \(\bar\theta\) denotes the mean of the estimates and \(\kappa\) is a constant whose value is insignificant for optimizing estimation strategies.
The optional factor \(n^{-1}\) in the definition of the error ellipsoid is used to counteract the shrinkage of error ellipsoid due to the increase in the number of samples, for this kind of benefit is not the focus of this work.
Enhancing estimation precision means shrinking the error ellipsoid as small as possible.
The QCRB implies that the error ellipsoid must lie outside the quantum-limited ellipsoid defined by~\cite{Helstrom1968}
\begin{equation} \label{eq:quantum_limited_ellipse}
    (v - \theta)^\TT \mathcal F (v - \theta) \leq \kappa^2.
\end{equation}
With measurement incompatibility, the error ellipsoid for any estimation strategy cannot stick close to the quantum-limited core simultaneously in all directions.
How to quantitatively describe the laws behind it is crucial to quantum multiparameter estimation and is answered by the TEUR for two-parameter estimation cases.

\begin{figure}
  \centering
  \includegraphics{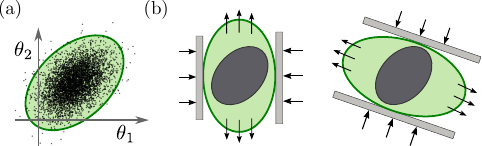}
  \caption{\label{fig:schematic}
    Schematic of error ellipse and its quantum-limited core.
    (a) Error ellipse represents the concentration characteristics of estimates in parameter space.
    (b) Measurement incompatibility forbids the error ellipse (filled with green) to stick close to the quantum-limited ellipse (filled with black) simultaneously in all directions.
  }
\end{figure}

For phase-space-displacement estimation, if the ancillary mode is prepared in a squeezed vacuum state with the squeeze parameter \(\zeta' = r'e^{i\varphi'}\), the extra error covariance in Eq.~\eqref{eq:example_error} is given by
\begin{equation}
  V(-Q', P') = \frac14 R \qty(-\frac{\varphi'}{2}) C(-2r') R\qty(-\frac{\varphi'}{2})^\TT.
\end{equation}
All the resultant covariance matrices saturate the TEUR, for any \(r'\) and \(\varphi'\).
\textit{For any error ellipse saturating the TEUR, it is impossible to squeeze it further in any direction without expanding it in other directions}, as illustrated in Fig.~\ref{fig:schematic}.

\begin{figure}
    \centering
    \includegraphics{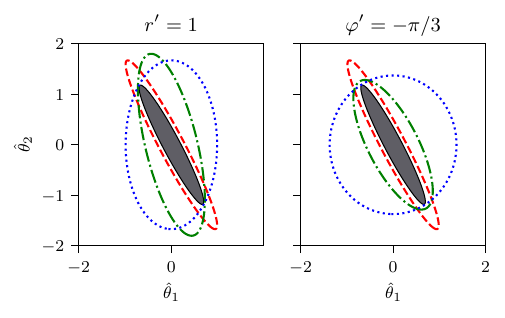}
    \caption{
      Error ellipses and quantum-limited ellipse (filled with black) for the joint estimation.
      Here, the system state is a displaced squeezed state with \(r=1\) and \(\varphi=\pi/3\).
      In the left panel, the measurements are given by \(\varphi'=-\pi/3\) (red dashed), \(\varphi'=0\) (green dash-dotted) and \(\varphi'=\pi/3\) (blue dotted), where \(r'\) is set to \(1\).
      In the right panel, the measurements are given by \(r'=1\) (red dashed), \(r'=0\) (green dash-dotted) and \(r'=-1\) (blue dotted), where \(\varphi'\) is set to \(-\pi/3\).
      Here, the true values of \(\theta_1\) and \(\theta_2\) are set to zero for convenience.
    }
    \label{fig:error_ellipse}
\end{figure}

Figure~\ref{fig:error_ellipse} plots the error ellipses for different measurements.
As shown in the left panel of Fig.~\ref{fig:error_ellipse}, the principal axes of the error ellipses are aligned with that of the quantum-limited ellipse, when \(\varphi'=-\varphi\).
The right panel of Fig.~\ref{fig:error_ellipse} shows that, by changing \(r'\), we can reduce the estimation error in one direction at the cost of increasing the estimation error in the other direction.
This cost will be tremendous if the error ellipse gets too close to the quantum-limited ellipse in any direction.

To rank the performance of the estimation strategies that saturate the TEUR, we need introduce other metrics.
For instance, the quantity \(|n \mathcal E|^{1/2}\) corresponds to the error volume/area~\cite{Xing2020}.
Using the Minkowski determinant inequality, we get the lower bound on the error area:
\begin{equation}
  |n \mathcal E|^{1/2} \geq |V(Q,P)|^{1/2} + |V(-Q',P)|^{1/2} = 1/2.
\end{equation}
Its minimum can be attained by \(r=r'\) and \(\varphi'=-\varphi\).

\vspace{8pt}
\noindent\textbf{\large Discussion\\}
To summarize, we have derived an estimation uncertainty relation, the TEUR, that captures the impact of measurement incompatibility and parameter correlation on quantum two-parameter estimation problems.
It has a simple form and is easy to evaluate; In comparison, the Holevo bound~\cite{Holevo2011}, which is more informative than the QCRB, is difficult to evaluate  despite recent progress~\cite{Albarelli2019,Sidhu2021}.
The TEUR completely describes the quantum limit of two-parameter estimation precision with pure states.
For cases of more than two parameters, it is promising to recast any multiparameter estimation problem into a series of two-parameter estimation problems through reparametrization.
At least for pure states, such an approach can always be implemented, as the quantum geometric tensor thereof can be transformed by reparametrization into block-diagonal forms with each block being at most two-dimensional~\cite{Hu2025,Wang2025a}.

The impact of the TEUR can be visualized by the error-ellipse approach.
This error ellipse is different from those often used in optical phase space for illustrating the uncertainties in the amplitudes of quadratures of the field for coherent states and squeezed states~\cite{Caves1980,Scully1997}.
The former reflects the actual estimation error for a given measurement while the latter is about properties of quantum states and corresponds to the quantum-limited ellipse in our phase-space complex displacement estimation example.
With the TEUR, we have treated the actual error ellipse of multiparameter estimation and the error ellipse in phase space for quantum optics in a unified framework.

Lastly, the TEUR would also be helpful to find the optimal measurement for single-parameter estimation.
Single parameter estimation can be embedded into a two-parameter one by introducing an auxiliary parameter, e.g., the phase-space real displacement estimation problem can be dilated to the joint estimation of the complex displacement.
When the auxiliary parameter is orthogonal to the original one and the incompatibility factor \(\gamma\) is unit, the TEUR Eq.~\eqref{eq:TEUR_extremal} leads us to a crucial clue about the optimal measurement for the original single-parameter estimation problem, that is, the optimal measurement must extract no distinguishability about the auxiliary parameter.
How to appropriately embed single-parameter estimation into a two-parameter one and construct the optimal measurement will be explored in future work.

\vspace{8pt}
\noindent\textbf{\large Methods\\}
The derivation of the TEUR is based on three key aspects: the information-regret-tradeoff relation (IRTR), the error ellipsoid representation, and parameter transformation.
We will explain these aspects below.\\

\noindent\textbf{IRTR.}
The IRTR establishes a fundamental tradeoff between the amount of classical Fisher information gained from measurements for different unknown parameters~\cite{Lu2021}.
Denote by \(F\) the classical Fisher information matrix (CFIM) defined as \(F_{\mu\nu}  = \sum_x [\partial_\mu p_\theta(x)] [\partial_\nu p_\theta(x)] / p_\theta(x) \), with which the classical Cram\'er-Rao bound reads \(\mathcal E \geq n^{-1} F^{-1}\).
The IRTR is given by
\begin{align} \label{eq:IRTR}
  \Delta_1^2 + \Delta_2^2 + 2 \sqrt{1 - c^2} \Delta_1 \Delta_2 \geq c^2,
\end{align}
where $\Delta_\mu := \sqrt{1 - F_{\mu\mu} / \mathcal{F}_{\mu\mu}}$ for \(\mu=1,2\) is the information regret ratio with respect to \(\theta_\mu\) and $c := \norm{\sqrt{\rho_\theta}[L_1, L_2]\sqrt{\rho_\theta}}_1 / (2 \sqrt{\mathcal{F}_{11}\mathcal{F}_{22}})$.
The IRTR is tight for pure state due to the mathematical properties of the Branciard inequality~\cite{Branciard2013}.
Note that only diagonal elements of the CFIM and the QFIM are involved in the IRTR, meaning that the IRTR does not account for the parameter correlations. \\

\noindent\textbf{Error ellipsoid/ellipse.}
An error ellipsoid defined as Eq.~\eqref{eq:error_ellipsoid} visualizes the concentration degree of parameter estimates from two perspectives.
First, many estimators, e.g., the maximum likelihood estimator~\cite{Fisher1922}, are asymptotically normally distributed, so that the error ellipsoid is given by a contour surface of the distribution and can be considered as a multidimensional generalization of confidence intervals with the confidence level \(\kappa\).
Second, the error ellipsoid with $\kappa = \sqrt{d+2}$ is known as the \emph{concentration ellipsoid}~\cite{Cramer1946, Cramer1946a}, over which the uniform distribution has the same covariance as the estimate distribution.

\begin{figure}[tb]
  \centering
  \includegraphics[]{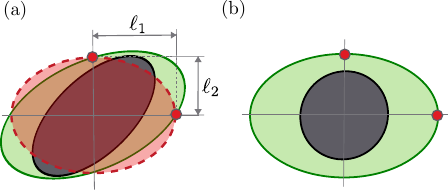}
  \caption{
    Error ellipse and its quantum limits.
    (a) For a general estimation problem, the error ellipse and the quantum-limited ellipse may be tilted due to parameter correlation.
    (b) Under a suitable reparametrization, the quantum-limited ellipse becomes a circle, while the error ellipse is rendered erect.
  }
  \label{fig:error_ellipse_quantum_limit}
\end{figure}

We here use error ellipse to analyze the impact of parameter correlation on the IRTR.
Parameter correlation in a quantum statistical model is manifested in the off-diagonal elements of the CFIM and the QFIM.
Due to the asymptotic attainability of the classical Cram\'er-Rao bound~\cite{Fisher1922,Cramer1946,Rao1945}, the smallest error ellipsoid in the asymptotic regime for a given measurement is the classical-limited ellipsoid defined as \((v - \theta)^\TT F (v - \theta) \leq \kappa^2\).
For convenience, we move the true value \(\theta\) to the origin of parameter space and set \(\kappa=1\).
Assuming that the error ellipse attains the Cram\'er-Rao bound, the intersection of the error ellipse with the horizontal and vertical axes, as shown in Fig.~\ref{fig:error_ellipse_quantum_limit}~(a), are given by \(\ell_1 = 1/\sqrt{F_{11}}\) and \(\ell_2 = 1/\sqrt{F_{22}}\).
The IRTR imposes a tradeoff between the diagonal elements of the CFIM and thereby restricts the simultaneous reduction of \(\ell_1\) and \(\ell_2\) for any quantum measurement.
However, the IRTR is irrelevant to the off-diagonal elements of the CFIM and the QFIM and thus does not account for the parameter correlation.
As a result, the IRTR does not give the full information about the permissible error ellipses.
For instance, the red dashed ellipse and the green solid ellipse in Fig.~\ref{fig:error_ellipse_quantum_limit}~(a) have the same intersections that satisfy the IRTR, but the former cannot be realized by any quantum estimation strategy as it breaks the quantum-limited ellipse.
\\

\noindent\textbf{Reparametrization.}
We use the reparametrization method to incorporate the parameter correlation into the a more complete tradeoff relation.
Specifically, we will introduce a special reparametrization, called the regular parametrization, to transform the quantum-limited ellipse into a circle and the error ellipse into an erect one, as shown in Fig.~\ref{fig:error_ellipse_quantum_limit}~(b).
This is the most effective way to apply the IRTR, as the CFIM and the QFIM are both diagonal.

Let \(\rho'_\phi = \rho_{\theta(\phi)}\) be a reparametrization given by a substitution \(\theta(\phi)\).
The CFIM and the QFIM with respect to the new vector parameter \(\phi\) can be obtained through \(F' = J^\TT F J\) and \(\mcF'= J^\TT \mcF J\), respectively, where \(J_{\mu\nu} = \pdv*{\theta_\mu}{\phi_\nu}\) is the Jacobi matrix.
Suppose that the Jacobi matrix of a specific reparametrization is \(J = \mcF^{-1/2} O\), where \(O\) is tentatively an arbitrary orthogonal matrix.
In such a case, the QFIM is changed into the identity matrix while the CFIM is changed into \(O^\TT \mcF^{-1/2} F \mcF^{-1/2} O\).
Furthermore, we can always find an orthogonal matrix \(O\) to diagonalize the CFIM and call such a reparametrization the regular parametrization, with which the CFIM is diagonal and the QFIM is the identity matrix.
In other words, the parameters in the regular parametrization are orthogonal~\cite{Cox1987}.

\vspace{8pt}
\noindent\textbf{Proof of the TEUR.}
We now outline the proof of the TEUR Eq.~\eqref{eq:Theorema_Egregium}. 
Applying the IRTR for the regular parametrization \(\rho'_\phi\), for which \(\mcF'_{\mu\nu} = \delta_{\mu\nu}\) and \(F'_{\mu\nu} = F'_{\mu\mu}\delta_{\mu\nu}\), it follows from the IRTR that
\begin{equation} \label{eq:proof_1}
  2 - \tr(F') + \sqrt{(1 - c'^2) \qty( 1 - \tr(F') + |F'| )} \geq c'^2,
\end{equation}
where \(c' := \norm{\sqrt{\rho'_\phi}[L'_1, L'_2]\sqrt{\rho'_\phi}}_1 / 2\) with \(L'_\mu\) being the symmetric logarithmic derivative operator with respect to \(\phi_\mu\).
In such a framework, all information about parameter correlation is encapsulated in the reparametrization process and can subsequently be incorporated into the tradeoff relation by reverting to the original parameters.

Next, we change the inequality Eq.~\eqref{eq:proof_1} back to the original parametrization.
Using \(  L'_\mu = \sum_\nu J_{\nu\mu} L_\nu\) and \(J=\mcF^{-1/2} O\), it can be shown that \(c'^2 = \gamma\), where \(\gamma\) is defined as Eq.~\eqref{eq:gamma}.
Substituting \(F' = O^\TT \mcF^{-1/2} F \mcF^{-1/2} O\) into Eq.~\eqref{eq:proof_1}, we have
\begin{align} \label{eq:proof_2}
  \qty(\sqrt{1- \gamma} + \sqrt{1 -\tr G + |G|})^2 \geq  |G|
\end{align}
with \(G := F \mcF^{-1}\).
Taking the square root of both sides of Eq.~\eqref{eq:proof_2} and using the identity \(\tr(G) = 1 + |G| - |I-G|\) and \(n \mcE \geq F^{-1}\), we obtain the TEUR given by Eq.~\eqref{eq:Theorema_Egregium} (see Supplementary Information for details).


\bibliography{ref.bib}

\vspace{8pt}
\noindent\textbf{Acknowledgements\\}
This work is supported by the Innovation Program for Quantum Science and Technology (Grant No.\ 2024ZD0301000) and the National Natural Science Foundation of China (Grants No.\ 92476118 and No.\ 12275062).

\vspace{8pt}
\noindent\textbf{Author contributions\\}
X.-M. L. supervised the project and derived the TEUR.
B.-S. H. conducted the analysis of the TEUR and the error ellipse approach for the phase-space estimation example and drafting the initial manuscript.
All authors contributed extensively to discussions throughout this work and collaboratively wrote the paper.

\noindent\textbf{Competing financial interests:}
The authors declare no competing financial interests.




\clearpage
\onecolumngrid 

\begin{center}
  \huge Supplementary Materials
\end{center}

\section{Derivation of the two-parameter estimation uncertainty relation} \label{sec:derivation}
We here give the detailed derivation of the two-parameter estimation uncertainty relation (TEUR).
We use the information regret tradeoff relation (IRTR)~\cite{Lu2021} as the starting point.
Let \(F\) and \(\mathcal F\) denote the classical Fisher information matrix (CFIM) and the quantum Fisher information matrix (QFIM), respectively.
The inefficiency of a quantum measurement for estimating the \(\mu\)-th parameter \(\theta_\mu\) can be quantified by the regret ratio defined by~\cite{Lu2021}
\begin{equation}
  \Delta_\mu \equiv \sqrt{\frac{\mathcal{F}_{\mu\mu} - F_{\mu\mu}}{\mathcal{F}_{\mu\mu}}}.
\end{equation}
The IRTR for any quantum measurement is given by
\begin{align} \label{seq:IRTR}
  \Delta_\mu^2 + \Delta_\nu^2 + 2 \sqrt{1 - c_{\mu\nu}^2} \Delta_\mu \Delta_\nu \geq c_{\mu\nu}^2
  \qq{with}
  c_{\mu\nu} \equiv \frac{\norm{\sqrt{\rho}[L_\mu,L_\nu]\sqrt{\rho}}_1}{2\sqrt{\mcF_{\mu\mu}\mcF_{kk}}},
\end{align}
where $\norm{X}_1 \equiv \tr\sqrt{X^{\dagger}X}$ denotes the Schatten-1 norm of an operator \(X\), \(L_\mu\) is the symmetric logarithmic derivative (SLD) operator about \(\theta_\mu\), and \(\rho\) is the density operator of the system.

Notice that the IRTR only involves the diagonal elements of the CFIM and thus gives no constraint on the off-diagonal elements of the CFIM.
We include the off-diagonal elements of the CFIM in the IRTR by utilizing parameter transformations in what follows.
Let $\phi=(\phi_1,\phi_2,\dots,\phi_d)$ be a set of new parameters, which are functions of the original parameters $\theta=(\theta_1,\theta_2,\dots, \theta_d)$.
The CFIM and the QFIM about \(\phi\) are given by~\cite{Liu2020}
\begin{equation} \label{seq:transform_rule}
  \mathcal F' = J^\TT \mathcal F J
  \qand
  F' = J^\TT F J,
\end{equation}
respectively, where \(J_{\mu\nu} \equiv \pdv*{\theta_\mu}{\phi_\nu}\) are the elements of the Jacobian matrix.
We henceforth assume that the transformation is invertible, i.e., the Jacobian matrix \(J\) is invertible.
We use the prime symbol \({}'\) to denote the quantities about the new parameters.
Let us now choose a special kind of parameter transformation whose Jacobian matrix is of the form \(\mathcal F^{-1/2} O\), where \(O\) is an arbitrary orthogonal matrix.
In such a case, according to Eq.~\eqref{seq:transform_rule}, the QFIM about \(\phi\) becomes the identity matrix no matter what the orthogonal matrix \(O\) is.
Meanwhile, the CFIM about \(\phi\) is given by
\begin{equation} \label{seq:F_prime}
  F' = O^\TT \mcF^{-1/2} F \mcF^{-1/2} O.
\end{equation}
Furthermore, we can always find an orthogonal matrix \(O\) to diagonalize the CFIM and call such parameters \(\phi\) \emph{the regular parameters}.
The regret ratio about the regular parameter \(\phi_\mu\) is \(\Delta'_\mu = \sqrt{1-F'_{\mu\mu}}\), as \(\mcF_{\mu\nu} = \delta_{\mu\nu}\).
Consequently, the IRTR about the two regular parameters becomes
\begin{equation} \label{seq:IRTR_regular}
  2 - F'_{\mu\mu} - F'_{\nu\nu} + 2\sqrt{\qty[1 - (c'_{\mu\nu})^2] (1-F'_{\mu\mu}) (1-F'_{\nu\nu})}
  \geq (c'_{\mu\nu})^2,
\end{equation}
where \(c'_{\mu\nu} = \frac12 \norm{\sqrt{\rho} [L'_\mu, L'_\nu] \sqrt{\rho} }_1\) with \(L'_\mu\) being the SLD operator about \(\phi_\mu\).
For two-parameter estimation problems, the IRTR about regular parameters can be expressed as
\begin{align} \label{seq:IRTR_two_parameter}
  2 - \tr(F') + 2\sqrt{\qty[1 - (c'_{12})^2] \qty[1 - \tr(F') + |F'|]}
  \geq (c'_{12})^2,
\end{align}
where we have used \(F'_{11} F'_{22} = |F'|\) as the CFIM about the regular parameters is diagonal.

We now transform the inequality Eq.~\eqref{seq:IRTR_two_parameter} from the regular parameters to the original ones.
To do so, we express \(c'_{12}\) with the quantities about the original parameters.
The SLD operators about \(\phi_\nu\) can be expressed as
\begin{equation}\label{seq:L_prime}
  L'_\nu = \sum_\mu J_{\mu\nu} L_\mu,
\end{equation}
which can be seen from \(\pdv*{\phi_\nu} = \sum_\mu J_{\mu\nu} \pdv*{\theta_\mu}\) and \(\pdv*{\rho}{\phi_\nu} = (L'_\nu \rho + \rho L'_\nu) / 2\).
It follows from Eq.~\eqref{seq:L_prime} that
\begin{align}
  [L'_1, L'_2] &= [J_{11} L_1 + J_{21} L_2, J_{12} L_1 + J_{22} L_2]  \nonumber \\
  &= (J_{11} J_{22} - J_{21} J_{12}) [L_1, L_2]  \nonumber \\
  &= |J| \, [L_1, L_2] \nonumber \\
  &= |\mcF|^{-1/2} [L_1, L_2],
\end{align}
where we have used \(|J| = |\mcF^{-1/2} O| = |\mcF|^{-1/2}\) in the last equality.
Therefore, we get
\begin{align} \label{seq:c_regular}
  c'_{12} &= \frac12 \norm{\sqrt{\rho} [L'_1, L'_2] \sqrt{\rho}}_1
  = \frac12 \norm{\sqrt{\rho} [L_1, L_2] \sqrt{\rho}}_1 \, |\mcF|^{-1/2}.
\end{align}
Meanwhile, it follows from Eq.~\eqref{seq:F_prime} that
\begin{equation} \label{seq:F_prime_to_F}
  \tr(F') = \tr(F \mcF^{-1})
  \qand
  |F'| = |F \mcF^{-1}|.
\end{equation}
Substituting Eqs.~\eqref{seq:c_regular} and \eqref{seq:F_prime_to_F} into Eq.~\eqref{seq:IRTR_two_parameter}, we get
\begin{equation} \label{seq:new_IRTR}
  \tr(F \mcF^{-1}) - 2 \sqrt{(1 - \gamma)\qty[1 - \tr(F \mcF^{-1}) + |F\mcF^{-1}|]}
  \leq 2-\gamma,
\end{equation}
where \(\gamma\) is the incompatibility factor defined as
\begin{equation}
  \gamma = \frac{ \norm{\sqrt{\rho} [L_1, L_2] \sqrt{\rho}}_1^2 }{4\,|\mcF|}.
\end{equation}

We now show how to rewrite Eq.~\eqref{seq:new_IRTR} regarding to the error-covariance matrix \(\mathcal E\).
For any two-dimensional matrix \(G\), it is easy to verify that
\begin{equation}
  \tr(G) = 1 + |G| - |I-G|.
\end{equation}
Taking \(G = F\mathcal{F}^{-1}\), it follows that
\begin{align}
  \tr(F \mathcal F^{-1}) = 1 + |F\mathcal{F}^{-1}| - |I-F\mathcal{F}^{-1}|.
\end{align}
Substituting the above expression into Eq.~\eqref{seq:new_IRTR} and multiplying both sides by \(|F^{-1}|\), we get
\begin{equation}
  |F^{-1}| + |\iqfim| - |F^{-1} - \iqfim| - 2\sqrt{(1 - \gamma)|F^{-1}| \times |F^{-1} - \iqfim|}
  \leq (2 - \gamma) |F^{-1}|,
\end{equation}
which can be simplified into
\begin{equation}
  (1 - \gamma) |F^{-1}| + |F^{-1} - \iqfim| + 2\sqrt{(1-\gamma)|F^{-1}| \times |F^{-1} - \iqfim|}
  \geq |\iqfim|.
\end{equation}
Taking the square root of both sides of the above inequality, we get
\begin{equation} \label{seq:eur_prior}
  \sqrt{(1 - \gamma)|F^{-1}|} + \sqrt{|F^{-1} - \iqfim|} \geq \sqrt{\qty|\iqfim|}.
\end{equation}
Notice that the classical and quantum Cram\'er-Rao bounds imply that \(|n \mathcal E| \geq |F^{-1}|\) and \(|n \mathcal E - \mathcal F^{-1}| \geq |F^{-1} - \mathcal F^{-1}|\).
Therefore, combining Eq.~\eqref{seq:eur_prior} and the classical Cram\'er-Rao bound, we get
\begin{equation}
  \sqrt{(1 - \gamma)|n\mcE|} + \sqrt{|n\mcE - \iqfim|} \geq \sqrt{\qty|\iqfim|}.
\end{equation}
Multiplying both sides of the above inequality by \(\sqrt{|\mathcal F|}\), we get
\begin{equation}
  \sqrt{\qty(1 - \gamma)|n\mcE\mcF|} + \sqrt{|n\mcE \mathcal F - I|} \geq 1,
\end{equation}
which is the TEUR presented in the main text.

\section{Brief review on the geometry of ellipse}\label{sec:ellipse}

\begin{figure}[bt]
  \centering
  \includegraphics[]{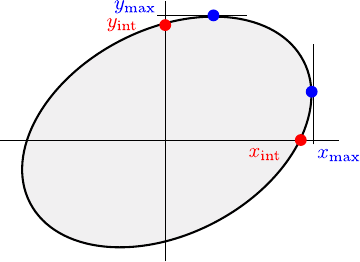}
  \caption{
    The intercepts and the maximum values of a tilted ellipse.
  }
  \label{sfig:ellipse}
\end{figure}

Let us consider an ellipse constituted by the coordinates \(v=(x,y)^\TT\) that satisfy \(v^\TT A\  v = 1\), where \(A\) is a real symmetric and positive two-dimensional matrix.
With the elements of \(A\) denoted by \(A_{jk}\), the equation of the ellipse can be expressed as
\begin{equation}\label{seq:ellipse}
  A_{11} x^2 + 2 A_{12} x y + A_{22} y^2 = 1
\end{equation}
The principal axes of the ellipse are given by the eigenvectors of $A$.
The lengths of the semi-axes are given by \(1/\sqrt{\lambda_j}\) for \(j=1,2\), where \(\lambda_j\) are the eigenvalues of $A$.
Figure~\ref{sfig:ellipse} illustrates the concepts of intercepts and maximum values of a tilted ellipse.
The intercepts of the ellipse are given by
\begin{equation}
  x_\mathrm{int} = 1 / \sqrt{A_{11}}
  \qand
  y_\mathrm{int} = 1 / \sqrt{A_{22}},
\end{equation}
which can be seen by setting \(y=0\) or \(x=0\) in the equation of the ellipse, that is, Eq.~\eqref{seq:ellipse}.
The maximum values of the ellipse are given by
\begin{equation}
  x_\mathrm{max} = \sqrt{(A^{-1})_{11}}
  \qand
  y_\mathrm{max} = \sqrt{(A^{-1})_{22}},
\end{equation}
which can be derived as follows.
To obtain the maximum values, e.g., \(y_\mathrm{max}\), we can express Eq.~\eqref{seq:ellipse} as
\begin{align}
  1 &= A_{11} \qty[ x + (A_{11})^{-1} A_{12}\ y]^2 + [A_{22} - (A_{11})^{-1} (A_{12})^2]\  y^2 \nonumber \\
  &= A_{11} \qty[ x + (A_{11})^{-1} A_{12}\  y]^2 + (A_{11})^{-1} |A| \ y^2,
\end{align}
where the symbol \(|A|\) denotes the determinant of the matrix \(A\).
It follows that
\begin{equation}
  y^2 \leq A_{11}/|A|,
\end{equation}
where the equality holds when \(x=-(A_{11})^{-1} A_{12} y\).
This means that
\begin{equation}
  y_\mathrm{max} = \sqrt{A_{11} / |A|}.
\end{equation}
Using the formula
\begin{equation} \label{seq:2d_inverse}
  \mqty(a & b \\ c & d)^{-1} = \frac1{ad-bc} \mqty(d & -b \\ -c & a).
\end{equation}
for the inverse of a two-dimensional matrix, we get \(y_\mathrm{max} = \sqrt{(A^{-1})_{22}}\).
The value of \(x_\mathrm{max}\) can be obtained in a similar way.

Substituting \(A\) by \((n\mcE)^{-1}\), we can get the intersections and the maximum values of the error ellipse.
Substituting \(A\) by \(F\) and \(\mcE\), we can get the intersections and the maximum values of the classical-limited ellipse and the quantum-limited ellipse, respectively.

\section{Example: Estimating a complex signal with squeezed state}\label{sec:example}

Let us consider the scenario of using a squeezed vacuum state as the initial state to estimate a complex parameter \(\alpha\) of the displacement operator \(D(\alpha) := \exp(\alpha a^\dagger - \alpha^* a)\) acting on the initial state.
The quantum state to be measured is the displaced squeezed state given by
\begin{equation}
  \ket{\alpha;\zeta}:=D(\alpha)S(\zeta)\ket{0},
\end{equation}
where \(S(\zeta):=\exp(\frac{\zeta^*}{2}a^2-\frac{\zeta}{2}(a^{\dagger})^2)\) is the squeeze operator with \(\zeta:=r\exp(i\varphi)\) being the squeeze parameter and \(\ket0\) is the vacuum state.
The parameters of interest here are the real and imaginary parts of $\alpha$, i.e., $\theta_1 = \Re\alpha$ and $\theta_2 = \Im\alpha$.

To apply the TEUR, we need to calculate the QFIM and the incompatibility factor;
Both of them can be determined by the quantum geometric tensor~\cite{Provost1980}:
\begin{equation}
  \mathcal{Q}_{jk} = \qty(\pdv{\bra{\alpha; \zeta}}{\theta_j}) \qty(\mathbbm{1} - \op{\alpha;\zeta}) \qty(\pdv{\ket{\alpha; \zeta}}{\theta_k}),
\end{equation}
where \(\mathbbm{1}\) denotes the identity operator.
The QFIM and the incompatibility factor for pure states are given by
\begin{equation}
  \mcF = 4 \Re\mcQ
  \qand
  \gamma = \frac{(\Im \mcQ_{12})^2}{|\Re \mcQ|},
\end{equation}
respectively.
Instead of straightforwardly calculating \(\mcF\) and \(\gamma\) from definitions, we here utilize the following relation between the displaced squeezed state and the squeezed coherent state~\cite[see Sec.~5.1]{Orszag2007}:
\begin{align} \label{seq:beta}
  D(\alpha)S(\zeta)\ket{0}
  = S(\zeta)D(\beta)\ket{0}
  \qq{with}
  \beta = \alpha\cosh r + \alpha^* e^{i\varphi} \sinh r.
\end{align}
Because \(S(\zeta)\) is a unitary operator independent of the parameters to be estimated, the quantum geometric tensor for the parametric states \(\ket{\alpha; \zeta}\) equals that for the coherent state \(D(\beta)\ket{0}\).
The quantum geometric tensor of \(D(\beta)\ket{0}\) with respect to the parameters \(\tilde\theta_1 := \Re\beta\) and \(\tilde\theta_2 = \Im\beta\) is known as~\cite{Lu2021}
\begin{equation}
  \tilde{\mcQ} = \mqty(1 & i \\ -i & 1).
\end{equation}
The quantum geometric tensor regarding to \(\theta_1\) and \(\theta_2\) can be obtained from \(\mathcal{Q} = J^\TT \tilde{\mcQ} J\), where \(J\) is the Jacobian matrix defined as $J_{\mu\nu}=\partial\tilde\theta_\mu / \partial \theta_\nu$.
It follows from Eq.~\eqref{seq:beta} that
\begin{equation}
  J = \mqty( \cosh r + \cos\varphi \sinh r &  \sin\varphi \sinh r \\
  \sin\varphi \sinh r &  \cosh r - \cos\varphi \sinh r),
\end{equation}
which can be decomposed into the combination of two rotation matrices and a scaling matrix, that is,
\begin{equation}
  J = R\qty(\frac{\varphi}{2}) C(r) R\qty(\frac{\varphi}{2})^\TT
  \qq{with} R(\varphi) = \mqty(\cos\varphi & - \sin\varphi \\ \sin\varphi & \cos\varphi )
  \qand
  C(r) = \mqty(\dmat[0]{e^r, e^{-r}}).
\end{equation}
We thus obtain the quantum geometric tensor with respect to the parameters \(\theta_1\) and \(\theta_2\):
\begin{align}
  \mathcal{Q} = J^\TT \tilde{\mcQ} J
  = R\qty(\frac{\varphi}{2}) C(2r) R\qty(\frac{\varphi}{2})^\TT + \mqty(0 & i \\ -i & 0),
\end{align}
which implies
\begin{equation} \label{seq:qfim}
  \mcF = 4 \,R\qty(\frac{\varphi}{2}) C(2r) R\qty(\frac{\varphi}{2})^\TT
  = 4 \mqty(
    \cosh 2r+\cos\varphi\sinh 2r & \sin \varphi\sinh 2r \\
  \sin \varphi\sinh 2r&\cosh 2r-\cos\varphi\sinh 2r )
  \qand
  \gamma = 1.
\end{equation}
Accordingly, the TEUR becomes
\begin{equation}\label{eq:new-inequality_gamma1}
  |\mcE \mcF - I| \geq 1.
\end{equation}

The quantum measurements that attain the TEUR can be constructed as follows.
Let us consider an ancillary mode whose annihilation operator are denote by $a'$.
Define the Hermitian operators for these two modes as
\begin{align}
  Q  &= \frac{a+a^{\dagger}}{2},   \quad P=\frac{a-a^{\dagger}}{2i},\\
  Q' &= \frac{a'+a'^{\dagger}}{2}, \quad P'=\frac{a'-a'^{\dagger}}{2i}.
\end{align}
Let us consider the following two observables
\begin{align}
  \mathcal A = Q - Q' \qand \mathcal B = P + P',
\end{align}
which can be jointly measured as \([\mathcal A, \mathcal B] = 0\).
We take the sample mean of the outcomes of \(\mathcal A\) and \(\mathcal B\) as the estimates for \(\theta_1\) and \(\theta_2\), respectively.
The unbiasedness condition requires that
\begin{equation} \label{seq:unbiasedness}
  \ev{\mathcal A} = \Re \alpha \qand \ev{\mathcal B} = \Im\alpha,
\end{equation}
where the expectation is taken with respect to the quantum state \(\ket{\alpha;\zeta}\otimes \ket{\mathrm{anc}}\) with \(\ket{\mathrm{anc}}\) denoting the initial state of the ancillary mode.
The unbiasedness condition Eq.~\eqref{seq:unbiasedness} implies that \(\ev{Q'}{\mathrm{anc}} = \ev{P'}{\mathrm{anc}} = 0\).
The error-covariance matrix for the above estimators is then given by \(\mcE = n^{-1} V(\mathcal A, \mathcal B)\), where we define the covariance matrix of operators \(X_\mu\) as
\begin{equation}
  [V(X_1,X_2)]_{\mu\nu}=\Re\ev{X_\mu X_\nu}-\ev{X_\mu}\ev{X_\nu}.
\end{equation}
Since the system and the ancilla are initially uncorrelated, we have
\begin{equation}
  V(\mathcal A, \mathcal B)
  = V(Q, P) + V(-Q', P').
\end{equation}

The state of the sensing system before measurement is the displaced squeezed state \(D(\alpha)S(\zeta)\ket{0}\).
The quantum expectation of an arbitrary operator \(X\) in the displaced squeezed state is equal to the quantum expectation of \(\widetilde X = S(\zeta)^\dagger D(\alpha)^\dagger X D(\alpha)S(\zeta)\) in the vacuum state.
It follows from
\begin{align}
  D(\alpha)^\dagger a D(\alpha) = a + \alpha
  \qand
  S(\zeta)^\dagger a e^{-i\varphi/2} S(\zeta) = a e^{-i\varphi/2} \cosh r - a^\dagger e^{i\varphi/2} \sinh r
\end{align}
that
\begin{align}
  \mqty(\widetilde a \\ \widetilde{a^\dagger})
  = \mqty(\dmat[0]{e^{i\varphi/2}}{e^{-i\varphi/2}})
  \mqty(\cosh r & - \sinh r \\ -\sinh r & \cosh r)
  \mqty(\dmat[0]{e^{-i\varphi/2}}{e^{i\varphi/2}})\mqty(a \\ a^\dagger) + \mqty(\alpha \\ \alpha^*).
\end{align}
The quadrature operators in the Heisenberg picture can be expressed as
\begin{equation}
  \mqty(\widetilde{Q} \\ \widetilde{P}) = K \mqty(\widetilde a \\ \widetilde{a^\dagger})
  \qq{with}
  K = \frac12 \mqty(1 & 1 \\ -i & i).
\end{equation}
It then follows that
\begin{align}
  \mqty(\widetilde{Q} \\ \widetilde{P})
  &= \underbrace{K \mqty(\dmat[0]{e^{i\varphi/2}}{e^{-i\varphi/2}}) K^{-1}}_{R \qty(\frac{\varphi}{2})}
  \underbrace{K \mqty(\cosh r & - \sinh r \\ -\sinh r & \cosh r) K^{-1}}_{C(-r)}
  \underbrace{K \mqty(\dmat[0]{e^{-i\varphi/2}}{e^{i\varphi/2}}) K^{-1}}_{R\qty(\frac{\varphi}{2})^\TT}
  K \mqty(a \\ a^\dagger)
  + K \mqty(\alpha \\ \alpha^*) \nonumber \\
  &= R \qty(\frac{\varphi}{2}) C(-r) R\qty(\frac{\varphi}{2})^\TT \mqty(Q \\ P)
  + \mqty(\Re\alpha \\ \Im\alpha). \label{seq:Q_P_transformation}
\end{align}
Using \(\ev{Q^2}{0} = \ev{P^2}{0} =1/4\) and \(\Re\!\ev{QP}{0} = \ev{Q}{0}=\ev{P}{0})=0\), we get
\begin{align} \label{seq:cov_system}
V(Q, P) &= R \qty(\frac{\varphi}{2}) C(-r) R\qty(\frac{\varphi}{2})^\TT
\mqty(\dmat[0]{\frac14}{\frac14})
R\qty(\frac{\varphi}{2}) C(-r) R\qty(\frac{\varphi}{2})^\TT \nonumber\\
&= \frac14 R \qty(\frac{\varphi}{2}) C(-2r) R\qty(\frac{\varphi}{2})^\TT.
\end{align}
Comparing Eq.~\eqref{seq:qfim} and Eq.~\eqref{seq:cov_system}, it can be seen that \(V(Q, P) = \mcF^{-1}\).
Therefore, the TEUR is equivalent to
\begin{equation}
|V(-Q', P')| \geq |\mcF^{-1}| = \frac{1}{16}.
\end{equation}
This means that the TEUR can be saturated if we prepare the ancillary mode in squeezed vacuum state.

Assume that the ancillary mode is prepared in a squeezed vacuum state \(\ket{\zeta'}\) with the squeeze parameter \(\zeta' = r' e^{i\varphi'}\).
It can be shown that the covariance matrix of the ancillary mode is given by
\begin{equation}
V(-Q', P') = \frac14 R \qty(-\frac{\varphi'}{2}) C(-2r') R\qty(-\frac{\varphi'}{2})^\TT.
\end{equation}
If we consider \(|V(\mathcal A, \mathcal B)|\) as a criterion, we have
\begin{align}
|V(\mathcal A, \mathcal B)|^{1/2} &= |V(Q, P) + V(-Q', P')|^{1/2} \\
&\geq |V(Q, P)|^{1/2} + |V(-Q', P')|^{1/2} \\
&= \frac1{4} |C(-2r)|^{1/2} + \frac1{4} |C(-2r')|^{1/2} \\
&= \frac{1}{2},
\end{align}
where the inequality follows from the Minkowski determinant inequality.
This bound can be saturated when \(\varphi'=-\varphi\) and \(r'=r\).

\end{document}